 \documentclass[final,5p,times,twocolumn]{elsarticle}

\usepackage{epsfig}
\usepackage{rotating}
\usepackage{float}
\restylefloat{table}
\usepackage{amssymb}
\usepackage{amsmath}
\usepackage{tabularx}

\journal{Physics Letters B}

\begin{document}

\begin{frontmatter}



\title{Nuclear aspects of neutral current non-standard $\nu$-nucleus reactions \\ 
       and the role of the exotic $\mu^-\to e^{-}$ transitions experimental limits}


\author[$^1$]{D.K. Papoulias}
\ead{dimpap@cc.uoi.gr}
\author[$^2$]{T.S. Kosmas}
\ead{hkosmas@uoi.gr}

\address{Division of Theoretical Physics, University of Ioannina, GR 45100 Ioannina, Greece}

\begin{abstract}
The nuclear aspects of flavour changing neutral current (FCNC) processes, predicted by various new-physics 
models to occur in the presence of nuclei, 
are examined by computing the relevant nuclear matrix elements within the context of the quasi-particle RPA
using realistic strong two-body forces. One of our aims is to explore the role of the non-standard interactions
(NSI) in the leptonic sector and specifically: 
(i) in lepton flavour violating (LFV) processes involving the neutral particles $\nu_\ell$ and $\tilde{\nu}_\ell$,
$\ell = e,\mu,\tau$ and
(ii) in charged lepton flavour violating (cLFV) processes involving the charged leptons $\ell^-$ or $\ell^+$.
As concrete nuclear systems we have chosen the stopping targets of $\mu^-\rightarrow e^-$ conversion experiments, 
i.e. the $^{48}\mathrm{Ti}$ nucleus of the PRIME/PRISM experiment at J-PARC and the $^{27}\mathrm{Al}$ 
of the COMET at J-PARC as well as of the Mu2e at Fermilab. These experiments have 
been designed to reduce the single event sensitivity down to $10^{-16}$--$10^{-18}$ in searching for charged 
lepton mixing events. Our goal is, by taking advantage of our detailed nuclear structure calculations and using 
the present limits or the sensitivity of the aforementioned exotic $\mu^- \rightarrow e^-$ experiments, to put 
stringent constraints on the parameters of NSI Lagrangians. 
\end{abstract}

\begin{keyword}

lepton flavour violation,  mu-to-e conversion, non-standard electroweak interactions, quasi-particle RPA, supernova neutrinos
\end{keyword}

\end{frontmatter}


\section{Introduction}
\label{Intro}

In recent years, ongoing extremely sensitive experiments searching for physics beyond the current Standard 
Model (SM) expect to see new physics or to set severe limits on various physical observables and particle 
model parameters \cite{Kuno,Molzon,Bernstein-Cooper}. In particular, current experiments searching for flavour changing neutral 
current (FCNC) processes in the leptonic sector \cite{Bernstein-Cooper,COMET,Mu2e-proposal,Mu2e,Bernstein,Kosm-talk,Valle} 
may provide insights and 
new results into the physics of charged lepton flavour violation (cLFV) \cite{Bernstein,Kosm-talk}, neutrino 
oscillation in propagation \cite{Valle} and others. The cLFV experiments, although they have not yet 
discovered any event, represent a very important probe to search for charged lepton mixing with significant 
implications on understanding various open issues in particle, nuclear physics and astrophysics 
\cite{Davidson,Barranco,Amanik_2005,Tomas-Valle-10}. To this purpose, exotic $\mu^-\to e^-$ conversion studies are 
interesting worldwide theoretically \cite{Kosm_PhysRep,Kos_Dep_Wal}
as well as experimentally with two experiments: 
(i) the COMET at J-PARC, Japan \cite{COMET}, and 
(ii) the Mu2e at Fermilab, USA \cite{Mu2e-proposal,Mu2e,Bernstein}. 
Both ambitious experiments expect to reach a single event sensitivity down to $10^{-16}$--$10^{-18}$.

The best previous limit for the $\mu^{-} \rightarrow e^{-}$ conversion was obtained by the 
SINDRUM-II collaboration at PSI on the reaction 
\begin{equation}
\mu^{-} + ^{48} \mathrm{Ti} \rightarrow e^{-} + ^{48} \mathrm{Ti} \, ,
\label{mue-conversion}
\end{equation}
as $R^{\mathrm{Ti}}_{\mu e} < 6.1 \times 10^{-13}$ \cite{Wintz} (many authors use the published upper limit $R^{\mathrm{Ti}}_{\mu e} < 4.3 \times 10^{-12}$ \cite{Dohmen}), where $R^{\mathrm{Ti}}_{\mu e}$ denotes the 
branching ratio of the   
$\mu^{-} \rightarrow e^{-}$ conversion rate divided by the total $\mu^{-}$-capture rate in the $^{48}$Ti nucleus. 
The COMET experiment, is expected to reach a high sensitivity, $R^{ \mathrm{Al}}_{\mu e} < 10^{-16}$ \cite{COMET} 
using $^{27}$Al as muon-stopping target while the Mu2e experiment aims to improve $R^{\mathrm{Al}}_{\mu e}$ even 
further, i.e. to a single event sensitivity $2\times 10^{-17}$, which with a background of 0.5 events will reach 
a target sensitivity $R^{\mathrm{Al}}_{\mu e} < 6\times 10^{-17}$ \cite{Mu2e-proposal,Mu2e,Bernstein}. 
The next decade experiments for cLFV, need very high intensity and quality muon beams, like those planed to be 
built at Fermilab for the Mu2e at Project-X and at J-PARC for the PRIME/PRISM experiments. The use of Project-X 
beams by the Mu2e experiment, expects to further decrease the upper bound to 
$R^{ \mathrm{Al}}_{\mu e} < 2 \times 10^{-18}$ \cite{mu2e-px}, while the PRIME experiment, based on 
the superior properties of the muon beam at J-PARC that can be delivered to the $^{48}\mathrm{Ti}$, 
may reach the sensitivity of $R^{ \mathrm{Ti}}_{\mu e} < 10^{-18}$ \cite{PRIME,Kuno-PRIME}.  

We should mention the most stringent upper 
bounds on purely leptonic cLFV processes presently available for $\mu-e$ transitions, 
namely, the new limit on the branching ratio of the $\mu^+ \rightarrow e^+ \gamma$ process, 
$Br(\mu^+ \rightarrow e^+ \gamma) < 5.7 \times 10^{-13}$, set very recently by the MEG experiment 
at PSI using one of the most intense continuous $\mu^+$ beams in the world \cite{MEG},  
and that of the $\mu \rightarrow e e e$ process set previously by the SINDRUM II collaboration 
in the value $Br(\mu^+ \rightarrow e^+ e^+ e^-) < 1.0 \times 10^{-12}$ \cite{mu-eee}. 

In recent works, neutral current (NC) neutrino scattering processes on leptons, nucleons and nuclei 
involving interactions that go beyond the SM (non-standard interactions, NSI, for short) have been 
examined \cite{Davidson,Barranco,Amanik_2005}. Such processes may be predicted from several extensions of the SM such as various realizations of the seesaw mechanism in the SM \cite{Kos_Dep_Wal,Schechter-Valle,Forero}, and left-right symmetric models \cite{Dep-Valle}. The reactions of this type that take place in nuclei are represented by 
\begin{equation}
\nu_{\alpha} (\tilde{\nu}_{\alpha}) + (A,Z) \rightarrow \nu_{\beta} (\tilde{\nu}_{\beta}) + (A,Z) \, ,
\label{neutrin-NSI}
\end{equation} 
($\alpha, \beta = e,\mu,\tau$)
and theoretically they can be studied under the same nuclear methods as the exotic cLFV 
process of $\mu^-\to e^-$ conversion in nuclei. Among the interesting 
applications of the reactions (\ref{neutrin-NSI}), those connected with the supernova 
physics may allow $\nu_e$ neutrinos to change flavour during core collapse creating $\nu_e$ 
neutrino holes in the electron-neutrino sea \cite{Amanik_2007} which may allow $e^{-}$-capture 
on nucleons and nuclei to occur and subsequently decrease the value of the electron fraction 
$Y_{e}$. Such non-standard interactions \cite{Friedland-solar,Friedland-atm2,Scholberg}
may suggest alterations in the mechanisms of neutrino-propagation through the supernova 
(SN) envelope and affect constraints put on the physics beyond the SM as well as on some scenarios 
of supernova explosion \cite{Barranco-Walle,Mir-Tort-Walle,Bar-Mir-Rashba}.
This motivated the investigation of the NSI in both LFV and cLFV processes in solar and supernova 
environment \cite{Kosm-A570,Kos_Kov_Schm} and motivated our present work too.
Furthermore, the impact of non-standard neutrino interactions on SN physics was the main motivation 
of works examining their effect on supernova when the neutrino self-interaction is taken into account
\cite{Tomas-Valle-10}. The extreme conditions under which neutrinos propagate after they are 
created in the SN core, may lead to strong matter effects. It is known that, in particular, the 
effect of small values of the NSI parameters can be dramatically enhanced in the inner strongly 
deleptonized regions \cite{Tomas-Valle-10}.

In general, low-energy astrophysical and laboratory neutrino searches provide crucial information 
towards understanding the fundamental electroweak interactions, within and beyond the SM. Well-known 
astrophysical neutrino sources like the solar, supernova, Geoneutrinos, etc., constitute 
excellent probes in searching for a plethora of neutrino physics applications and new-physics open issues 
\cite{Kosm-Oset}. Since neutrinos interact extremely weakly
with matter, they may travel astronomical distances and reach the Earth \cite{SK,SNO,KamLAND}, etc. 
The recorded $\nu$-signals in sensitive terrestrial nuclear detectors of low-energy neutrinos
\cite{Hirata-Bionta,Keil}, could be simulated providing useful information relevant to the 
evolution of distant stars, the core collapse supernovae, explosive nucleosynthesis \cite{Haxton}, 
neutrino oscillation effects and others. 
Recently it became feasible to detect neutrinos by exploiting the NC interactions and measuring 
the nuclear recoil signal by employing detectors with very low-threshold energies \cite{pion-DAR-nu,Louis}. 
The NC interactions, through their vector components can lead to an additive contribution (coherence) 
of all nucleons in the target nucleus 
\cite{Horowitz,Biassoni,Freedman,Giom-Vergados,Monroe_Fischer,Don-Wal}.

The main purpose of the present Letter is to explore the nuclear physics aspects of the $\nu$-nucleus 
reactions of Eq. (\ref{neutrin-NSI}) focusing on the role of the NSI which have not been studied in 
detail up to now. We should stress that, our strategy in studying the nuclear aspects of FCNC in nuclei,  
is to carry out realistic cross sections calculations for the exotic processes (\ref{mue-conversion}) 
and (\ref{neutrin-NSI}), including NSI terms in the relevant effective Lagrangian. The required nuclear 
matrix elements are evaluated within the context of the quasi-particle RPA, considering both coherent 
and incoherent processes by applying the advantageous state-by-state method developed in Refs. 
\cite{Kosm-A570,vtsak-tsk-1,vtsak-tsk-2}. As a first step, we perform calculations 
for $gs\to gs$ 
transitions of the reactions (\ref{neutrin-NSI}) by solving the BCS equations, for even-even nuclear systems, 
and employing the experimental nuclear charge densities \cite{deVries} for odd-$A$ nuclei. For comparison 
of our results with those of other methods \cite{Barranco,Amanik_2005,Amanik_2007,Horowitz,Biassoni}, SM 
cross sections calculations are also carried out. More specifically, our present results refer to the even-even 
$^{48}\mathrm{Ti}$ isotope, the stopping target of SINDRUM II and PRIME/PRISM $\mu^-\to e^-$ experiments. 
We perform similar calculations for processes (\ref{neutrin-NSI}) in the $^{27}\mathrm{Al}$ nucleus proposed 
as detector material in Mu2e and COMET experiments. Finally, we will use the experimental upper limits 
of the cLFV processes to put robust bounds on model parameters of the relevant Lagrangians and the 
ratios of the NSI contributions with respect to the SM ones.

\section{Description of the formalism}
\label{chapt2}

The non-standard $\nu$-nucleus processes (\ref{neutrin-NSI}) and the exotic cLFV $\mu^-\to e^-$ conversion 
in nuclei \cite{Kuno,Kosm_PhysRep,Kos_Dep_Wal,Kos_Kov_Schm}, can be predicted within the
aforementioned new-physics models \cite{Kos_Dep_Wal}.
%
\begin{figure}[t]
\begin{center}
\includegraphics[width=0.45\textwidth]{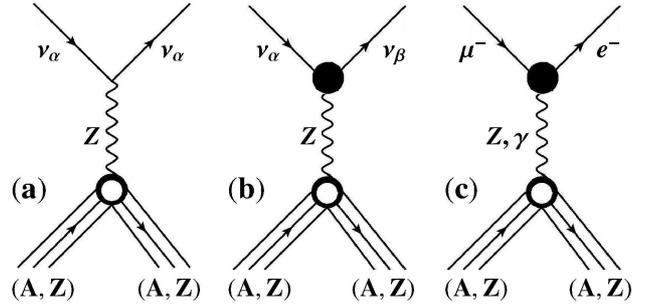}
\end{center}
\caption{ Nuclear level Feynman diagrams for: \textit{(a)} SM Z-exchange neutral current $\nu$-nucleus 
reactions, \textit{(b)} non-standard Z-exchange $\nu$-nucleus reactions, and \textit{(c)} Z-exchange 
and photon-exchange $\mu^{-}\rightarrow e^{-}$ in the presence of a nucleus (muon-to-electron conversion). 
The non-standard (cLFV or LFV) physics enters in the complicated vertex denoted by the bullet $\bullet$.} 
\label{fig.1}
\end{figure}
%
In Fig. \ref{fig.1} we show some nuclear-level Feynman diagrams representing the exchange of a $Z$-boson 
between a lepton and a nucleon for the cases of $\nu$-nucleus scattering in the SM (Fig. \ref{fig.1}(a)) 
and in the non-standard interactions of neutrinos with nuclei (Fig. \ref{fig.1}(b)). We also show the 
exchange of a $Z$-boson or a $\gamma$-photon in the $\mu^-\to e^-$ conversion, Fig. \ref{fig.1}(c)
\cite{Kosm_PhysRep,Kos_Dep_Wal}. The leptonic vertex in the cases of Fig. \ref{fig.1}(b),(c) is a 
complicated one. A general effective Lagrangian that involves SM interactions ($\mathcal{L}_{\mathrm{SM}}$) 
and NSI ($\mathcal{L}_{\mathrm{NSI}}$) with a non-universal (NU) term and a flavour changing (FC) term can 
be written as
\begin{equation}
\mathcal{L}_{\mathrm{tot}} = \mathcal{L}_{\mathrm{SM}} + \mathcal{L}_{\mathrm{NSI}} = 
     \mathcal{L}_{\mathrm{SM}} + \mathcal{L}_{\mathrm{NU}} + \mathcal{L}_{\mathrm{FC}} \, . 
\label{tot_Lagr}
\end{equation}
The individual components $\mathcal{L}_{\mathrm{SM}}$ and $\mathcal{L}_{\mathrm{NSI}}$ of this Lagrangian are explained in the next subsections.

For a concrete example, it has been proposed \cite{Schechter-Valle} that, even small deviation from unitary lepton mixing matrix, may cause sizeable NSI effects and potentially large LFV \cite{Forero}. The non-trivial structure of electroweak currents in low-scale seesaw Lagrangians leads to non-unitary lepton mixing matrix $N_{\alpha \beta}$, which can be parametrized as $N\equiv(1- n)U$. $U_{\alpha \beta}$ is a unitary matrix and $n_{\alpha \beta}$ a model depended non-standard matrix ($\alpha, \beta = e,\mu,\tau$) which takes specific form within seesaw mechanisms \cite{Forero}.  
\subsection{Non-standard $\nu$-nucleus reaction cross sections}
The neutral current non-standard neutrino interactions addressed here, are described by a quark-level Lagrangian, 
$\mathcal{L}_{\mathrm{NSI}}$, parametrized (for energies $\ll M_{Z}$) as \cite{Barranco,Scholberg,Barranco-Walle} 
\begin{equation}
\mathcal{L}_{\mathrm{NSI}} = - 2\sqrt{2} G_F \sum_{\begin{subarray}{c} f= \, u,d\\ \alpha,\beta = \, e,\mu,\tau\end{subarray}} 
\epsilon_{\alpha \beta}^{f P}\left[\bar{\nu}_{\alpha}\gamma_\rho L \nu_\beta\right]\left[\bar{f}\gamma^\rho P f\right],
\label{NSI_Lagr}
\end{equation}
where three light neutrinos $\nu_{\alpha}$ with Majorana masses are considered, $f$ denotes a first generation 
SM quark and $P=\lbrace L, R\rbrace$ are the chiral projectors. The Lagrangian (\ref{NSI_Lagr}) contains flavour 
preserving non-SM terms, known as non-universal (NU) interactions 
that are proportional to $\epsilon_{\alpha \alpha}^{f P}$, as well as flavour-changing (FC) terms proportional 
to $\epsilon_{\alpha \beta}^{f P}$, $\alpha\neq\beta$. These couplings are taken with respect to the strength 
of the Fermi coupling constant $G_F$ \cite{Barranco,Barranco-Walle}. For the polar-vector couplings we are mainly  
interested 
in the present work, it holds $\epsilon_{\alpha\beta}^{f V}=\epsilon_{\alpha\beta}^{f L} + \epsilon_{\alpha\beta}^{f R}$, 
while for the axial-vector couplings $\epsilon_{\alpha\beta}^{f A}=\epsilon_{\alpha \beta}^{f L} - \epsilon_{\alpha\beta}^{f R}$. 

The nuclear physics aspects of the non-standard $\nu$-matter reactions can be studied by transforming the 
Lagrangian (\ref{NSI_Lagr}) to the nuclear level where the hadronic current is written in terms of NC nucleon 
form factors (functions of the four momentum transfer) \cite{Kos_Kov_Schm}. In the general case of the inelastic 
scattering of neutrinos on nuclei, the magnitude of the three momentum transfer, $q = \vert{\bf q}\vert$, obtained 
from the kinematics of the reaction, is a function of the scattering angle of the outgoing neutrino $\theta$ 
(laboratory frame), the initial, $E_{i}$, and final, $E_{f}$, neutrino energies, 
as well as the excitation energy of the target nucleus $\omega$ as,
$q^2=\omega^2+ 2 E_{i} E_{f} \left(1 - \cos \theta \right)$ \cite{Don-Wal,vtsak-tsk-1}.
In the special case of the coherent (elastic) channel we focus in this work ($\omega=0$ and $E_i = E_f \equiv E_\nu$), 
only $gs \rightarrow gs$ transitions occur (for spin-zero nuclei) 
and we have $q^2 = 2 E_\nu^2 (1-\cos \theta)$ or $q = 2 E_\nu \sin (\theta/2)$.  

The coherent differential cross section with respect to the scattering angle $\theta$ for NSI $\nu$-nucleus 
processes is written as 
\begin{equation}
\frac{d \sigma_{\mathrm{NSI},\nu_{\alpha}}}{d \cos \theta} = \frac{G_{F}^{2}}{2 \pi} E_{\nu}^{2} \left(1 + 
\cos \theta \right)\left\vert 
\langle gs \vert \vert G_{V,\nu_{\alpha}}^{\mathrm{NSI}}(q) \vert \vert gs \rangle \right \vert ^{2},
\label{NSI_dcostheta}
\end{equation}
($\alpha = e,\mu,\tau$, denotes the flavour of incident neutrinos) where $\vert gs \rangle$ represents 
the nuclear ground state (for even-even nuclei, like 
the $^{48}\mathrm{Ti}$, $\vert gs \rangle=\vert J^\pi \rangle\equiv\vert 0^+ \rangle$). The nuclear matrix element, 
that arises from the Lagrangian (\ref{NSI_Lagr}), takes the form 
\begin{equation}
\begin{aligned}
& \left\vert {\cal M}^{\mathrm{NSI}}_{V,\nu_{\alpha}} \right \vert ^{2} \equiv
\left\vert \langle gs \vert \vert G_{V,\nu_{\alpha}}^{\mathrm{NSI}}(q)  \vert \vert gs \rangle \right \vert ^{2}  = \\  & 
 \left[ \left( 2 \epsilon_{\alpha \alpha}^{uV} + \epsilon_{\alpha \alpha}^{dV} \right) Z F_Z (q^2) +  
\left( \epsilon_{\alpha\alpha}^{uV} + 2\epsilon_{\alpha\alpha}^{dV} \right) N F_N (q^2) \right]^2 \\
&  + \sum_{\beta \neq \alpha} \left[\left( 2 \epsilon_{\alpha \beta}^{uV}+ \epsilon_{\alpha \beta}^{dV} 
\right) Z F_Z (q^2)+ \left(\epsilon_{\alpha \beta}^{uV}+ 2 \epsilon_{\alpha\beta}^{dV} \right)N F_N (q^2)\right]^2,
\end{aligned}
\label{GV}
\end{equation}
($\beta = e,\mu,\tau$) where $F_{Z(N)}$ denote the nuclear (electromagnetic) form factors for protons 
(neutrons) entered due to the CVC theory. We note that in the adopted NSI model, the coherent NC $\nu$-nucleus 
reaction is not a flavour blind process. By considering the nuclear structure details, the cross sections provided 
by Eq. (\ref{NSI_dcostheta}), become more realistic and accurate \cite{Scholberg} (in Ref. \cite{Barranco} 
the variation versus the momentum transfer of the nuclear form factor is neglected, which for supernova neutrino 
studies is a rather crude approximation \cite{Pap-Kosm-NPA}). 

From an experimental physics point of view, many neutrino detectors are more sensitive to the recoil energy 
of the nuclear target, $T_N$, than to the scattering angles, $\theta$. Therefore, it is also important 
to compute the differential cross sections $d\sigma/dT_N$. For coherent scattering the nucleus 
recoils (intrinsically it remains unchanged) with energy which, in the approximation $T_{N} \ll E_{\nu}$ 
(low-energy limit), is maximized as, $T_N^{\text{max}}=2 E_\nu^2/(M+ 2 E_\nu)$, with $M$ being the nuclear mass 
\cite{Giom-Vergados,Monroe_Fischer}. Then, to a good approximation, the square of the three momentum transfer, 
is equal to $q^2 = 2 M T_N$, and the coherent NSI differential cross section with respect to $T_{N}$ is written 
as
\begin{equation}
\frac{d\sigma_{\mathrm{NSI},\nu_{\alpha}}}{dT_N} = \frac{G_F^2 \,M}{\pi} \left(1- 
\frac{M\,T_N}{2 E_\nu^2}\right)\left\vert\langle gs\vert\vert 
G_{V,\nu_{\alpha}}^{\mathrm{NSI}} (q) \vert\vert gs \rangle \right \vert ^{2}\, .
\label{NSI_dT}
\end{equation}
Both Eqs. (\ref{NSI_dcostheta}) and (\ref{NSI_dT}) are useful for studying the nuclear physics of NSI of 
neutrinos with matter.
 
Furthermore, by performing numerical integrations in Eq. (\ref{NSI_dcostheta}) over the scattering 
angle $\theta$ or in Eq. (\ref{NSI_dT}) over the recoil energy $T_N$, one can obtain integrated (total) 
coherent NSI cross sections, $\sigma_{\mathrm{NSI},\nu_\alpha}$. The 
individual cross sections $\sigma_{\mathrm{NU,\nu_{\alpha}}}$ and $\sigma_{\mathrm{FC,\nu_{\alpha}}}$ 
may be evaluated accordingly \cite{Pap-Kosm-NPA}. 
\subsection{SM coherent $\nu$-nucleus cross sections}
At low and intermediate neutrino energies considered in this Letter, the effective (quark-level) SM 
$\nu$-nucleus interaction Lagrangian, $\mathcal{L}_{\mathrm{SM}}$, reads 
\begin{equation}
\mathcal{L}_{\mathrm{SM}} = - 2 \sqrt{2} G_{F} \sum_{ \begin{subarray}{c} f= \, u,d \\ \alpha= e, \mu,\tau 
\end{subarray}} g_P^f   
\left[ \bar{\nu}_{\alpha} \gamma_{\rho} L \nu_{\alpha} \right] \left[ \bar{f} \gamma^{\rho} P f \right],
\label{SM_Lagr}
\end{equation}
where, $g_f^P$ are the $P$-handed SM couplings of $f$-quarks ($f=u,d$) to the $Z$-boson. We mention that, compared to previous studies \cite{Amanik_2005,Amanik_2007}, 
we have taken into consideration the $\nu-u$ quark interaction [see Eq. (\ref{GV})], in addition to the 
momentum dependence of the nuclear form factors. 

For coherent $\nu$-nucleus scattering, the SM angle-differential cross section is given from an expression 
similar to Eq. (\ref{NSI_dcostheta}) with the nuclear matrix element being that of the Coulomb operator 
$\hat{\mathcal{M}}_0(q)$ (product of the zero-order spherical Bessel function times the zero-order 
spherical harmonic \cite{Don-Wal}). This corresponding matrix element can be cast in the form \cite{Kosm-A570}
\begin{equation}
\left\vert {\cal M}^{\mathrm{SM}}_{V,\nu_{\alpha}} \right \vert ^{2} \, \equiv \,
\left\vert\langle gs \vert\vert \hat{\mathcal{M}}_0 \vert\vert gs \rangle\right \vert^2 = 
\left[g^p_V Z F_Z (q^2) + g^n_V N F_N (q^2) \right]^2 ,
\label{SM-ME}
\end{equation} 
where, $g^{p(n)}_V$ is the known  polar-vector coupling of proton (neutron) to the $Z$ boson (see 
Fig. \ref{fig.1}(a)). In the low energy limit, one can also write in a straightforward manner the corresponding differential cross section with respect to the nuclear recoil energy, $T_N$ \cite{Giom-Vergados,Monroe_Fischer}). 
In this work, starting from original differential cross sections $d\sigma_{\lambda,\nu_\alpha}/d\cos\theta$ 
and $d\sigma_{\lambda,\nu_{\alpha}}/dT_N$, we evaluated individual angle-integrated cross sections of the 
form $\sigma_{\lambda,\nu_\alpha} (E_\nu)$, with $\alpha = e,\mu,\tau$, and 
$\lambda= \mathrm{tot, SM, NU, FP, FC}$, where under FC, the six processes
$\nu_e\leftrightarrow\nu_\mu, \, \nu_e\leftrightarrow\nu_\tau, \, \nu_\mu\leftrightarrow\nu_\tau$ are included (obviously, $\sigma_{\nu_\alpha\rightarrow\nu_\beta}=\sigma_{\nu_\beta\rightarrow \nu_\alpha}$) for both nuclei, $^{48}\mathrm{Ti}$ and $^{27}\mathrm{Al}$. A great part of these results is presented and used to compute folded cross sections below (for more results see Ref. \cite{Pap-Kosm-NPA}).

\section{Results and discussion}
\subsection{Nuclear Structure calculations}
At first, we studied the nuclear structure details of the matrix elements entering 
Eqs. (\ref{NSI_dcostheta})-(\ref{NSI_dT}) and Eq. (\ref{SM-ME}) that reflect the 
dependence of the 
coherent cross section on the incident $\nu$-energy $E_{\nu}$ and the scattering angle $\theta$ (or 
the recoil energy $T_{N}$). For the even-even $^{48}\mathrm{Ti}$ nucleus, the stopping target of the 
PSI \cite{Wintz,Dohmen} and PRIME \cite{PRIME,Kuno-PRIME} experiments, this study involves realistic 
nuclear structure calculations for the cross sections 
$d\sigma_{\lambda,\nu_{\alpha}}/d\cos\theta$ and $d\sigma_{\lambda,\nu_{\alpha}}/dT_N$, performed after
constructing the nuclear ground state $\vert gs\rangle$ by solving iteratively the BCS equations 
\cite{vtsak-tsk-1}. Then, the nuclear form factors for protons (neutrons) are obtained as \cite{Kosm-A570}
\begin{equation}
F_{N_n}(q^2) = \frac{1}{N_n}\sum_j [j]\, \langle j\vert j_0(qr)\vert j\rangle\left(\upsilon^j_{N_n}\right)^2
\end{equation}
with $[j]=\sqrt{2 j +1}$, $N_{n}=Z \,\,(\mathrm{or}\,\, N)$. $\upsilon^j_{N_{n}}$ denotes the occupation 
probability amplitude of the $j$-th single nucleon level. The chosen active model space consists of the 
lowest 15 single-particle $j$-orbits, $j \equiv (n , \ell, 1/2)j$ without core, up to major h.o. quanta $N=4 \hbar \omega$. The required monopole (pairing) 
residual interaction, obtained from a Bonn C-D two-body potential was slightly 
renormalized with the two parameters $g^{p,n}_{\mathrm{pair}}$ ($g^{p}_{\mathrm{pair}}=1.056$, for proton 
pairs, and $g^{n}_{\mathrm{pair}}=0.999$, for neutron pairs).

We note that, we have devoted a special effort on the accurate construction of the nuclear ground state, (i) because the coherent channel is the dominant one for 
the neutral current SM $\nu$-nucleus processes and we assumed that this holds also for NSI processes, and (ii) because 
in a next step we are intended to perform extensive incoherent cross sections calculations where all 
accessible final nuclear states will be built on the present ground state. 

For the odd-$A$ $^{27}\mathrm{Al}$ nucleus (its ground state spin is 
$\vert gs\rangle =\vert J^\pi\rangle =\vert (5/2)^+\rangle$), the stopping target of Mu2e and COMET 
experiments, we obtained the form factor $F_{Z}(q^{2})$, through a model independent 
analysis (using a Fourier-Bessel expansion model) of the electron scattering data for the charge 
density distribution of this isotope \cite{deVries}. Since similar data for $F_{N}(q^{2})$
$^{27}\mathrm{Al}$ are not available, we considered (to a rather satisfactory approximation) that 
$F_{N} \simeq F_{Z}$ (a difference up to about $10 \%$ usually appears for medium and heavy 
nuclear systems \cite{deVries}). 
The momentum dependence of the nuclear form factors was ignored by some authors 
\cite{Barranco} which at low $\nu$-energies relevant for solar neutrinos is practically a good 
approximation, but for energies relevant to supernova neutrinos addressed in this work, it may 
lead to differences of even an order of magnitude \cite{Pap-Kosm-NPA}. 
\subsection{ Integrated coherent $\nu$-nucleus cross sections}

\begin{table*}[ht]
\centering
\begin{tabular}{{c|cccccccc}}
\hline \hline   \\
 $\nu_{\alpha}$ & $(A,Z)$ & $R_{\mathrm{tot}}$ & $R_{\mathrm{NU}}$ & $R_{\mathrm{FP}}$ & 
$R_{\nu_{\alpha} \leftrightarrow \nu_{e}}$ & $R_{\nu_{\alpha} \leftrightarrow \nu_{\mu}}$ & 
$R_{\nu_{\alpha} \leftrightarrow \nu_{\tau}}$ & \\
 \hline  
  & $^{48}\mathrm{Ti}$ & 1.037 & 0.002 & 0.905  & -  & $0.121 \times 10^{-4}$  & 0.130  & \\[-1ex]
\raisebox{1.5ex}{ $\nu_{e}$ } & $^{27}\mathrm{Al}$ & 1.044 & 0.003  & 0.902 & - & $0.130 \times 10^{-4}$ & 0.139 & \\
\hline 
 & $^{48}\mathrm{Ti}$ & 1.293 & 0.001  & 0.929  & $0.121 \times 10^{-4}$  & -  & 0.361 & \\[-1ex]
\raisebox{1.5ex}{ $\nu_{\mu}$ } & $ ^{27}\mathrm{Al}$ & 1.318 & 0.001 & 0.927 & $0.130 \times 10^{-4}$ & - & 0.387 & \\
 \hline \hline
\end{tabular}
\caption{ The ratios $R_{\lambda,\nu_{\alpha}}$ (for the definition see Eq. (\ref{R-lamb-alp}) 
in the text) of all possible $\nu_{\alpha} + (A,Z) \rightarrow\nu_{\beta} + (A,Z) $ processes. 
They have been evaluated in their assymptotic values reached at  $E_{\nu} \approx 120 \, \mathrm{MeV}$.}
\label{table1}
\end{table*}

In the next step of our calculational procedure we obtained angle-integrated coherent $\nu$-nucleus 
cross sections by integrating numerically Eq. (\ref{NSI_dcostheta}) over angles [or Eq. (\ref{NSI_dT}) 
over $T_N$] for the various interaction components as
\begin{equation}
\sigma_{\lambda,\nu_{\alpha}}(E_\nu) = \int\frac{d\sigma_{\lambda,\nu_{\alpha}}}{d\cos \theta}(\theta , 
E_\nu) \,\, d\cos \theta  \, , 
\end{equation} 
($\lambda= \mathrm{tot, SM, NU, FP, FC }$). We found that the exotic FCNC processes $\nu_\alpha\rightarrow \nu_\beta$ 
in $^{48}\mathrm{Ti}$ have significantly lower cross section compared to the SM one. From the obtained FCNC $\nu$-nucleus cross sections the most challenging result corresponds to the 
$\nu_{\mu} \rightarrow \nu_{e}$ transition (and to its lepton conjugate process, $\nu_e\to\nu_\mu$). 
This is mainly due to the severe constraint $\epsilon_{\mu e}^{f P} = 2.9 \times 10^{-4}$  inserted in 
the Lagrangian (\ref{NSI_Lagr}) which has been derived from the nuclear $\mu^-\to e^-$ 
conversion experimental limits on cLFV branching ratio \cite{COMET,Mu2e-proposal,Mu2e,Bernstein}. 
We remind that, in this work we have employed the NSI parameters $\epsilon_{\alpha \beta}^{f V}$ 
(except the $\epsilon_{\mu e}^{f V}$) derived from various experimental 
bounds in Ref. \cite{Davidson}. 

By exploiting our cross sections $\sigma_{\lambda,\nu_{\alpha}}(E_\nu)$, we find it interesting 
to estimate the ratio of each of the individual cross sections, $\sigma_{\lambda,\nu_\alpha}$, with 
respect to the SM cross sections defined as
\begin{equation}
R_{\lambda,\nu_{\alpha}}(E_{\nu}) = \frac{\sigma_{\lambda,\nu_{\alpha}}(E_{\nu})}{\sigma_{\mathrm{SM}}(E_{\nu})}
 , \qquad \lambda= \mathrm{tot, NU, FP, FC} \, . 
\label{R-lamb-alp} 
\end{equation}
For $^{48}\mathrm{Ti}$, the latter ratios initially are slowly increasing functions of $E_\nu$, 
but eventually (for energies higher than about $80-120$ MeV) they tend asymptotically to 
the values listed in Table \ref{table2}. For $^{27}\mathrm{Al}$, however, the ratios 
$R_{\lambda,\nu_{\alpha}}$ are energy independent which is a consequence of the different treatment applied in studying the nuclear structure details than that followed for  $^{48}\mathrm{Ti}$. From the comparison of the results of Table \ref{table2} with those 
of the method \cite{Barranco}, we conclude
that our realistic calculations are important in the case of $^{48}\mathrm{Ti}$ nucleus, 
where the BCS method gave us $F_N \neq F_Z$ and, hence, the results obtained for $R_{\lambda,\nu_{\alpha}}$ 
differ from those given by Ref. \cite{Barranco}. For $^{27}\mathrm{Al}$, however, for 
which we considered $F_{N} \simeq F_{Z}$, the dependence on the nuclear structure parameters in the numerator 
and denominator of Eq. (\ref{R-lamb-alp}) cancel out and, then, our predictions for $R_{\lambda,\nu_{\alpha}}$ 
are equal to those of Ref. \cite{Barranco}.

It is worth noting that, some constraints coming from solar \cite{Friedland-solar} and atmospheric 
\cite{Friedland-atm2} neutrino data indicate that the NSI might be large, while according to the 
present experimental data, $\epsilon_{\tau \tau}^{f V}$ is unacceptably large and, consequently, it derives  
unrealistic results (the corresponding FP and NU cross sections, not included here, are larger than the SM ones)
\cite{Davidson,Scholberg}.
\subsection{Supernova neutrino fluxes and expected event rates}
One of the most interesting connections of our present calculations with ongoing and future neutrino 
experiments is related to 
supernova $\nu$-detection. As it is known, in SN explosions most of the energy is 
released by $\nu$-emission. Then, the total neutrino flux, $\Phi(E_\nu)$, arriving at a terrestrial detector reads \cite{Horowitz,Biassoni}
\begin{equation}
\Phi(E_\nu) = \sum_{\alpha} \Phi_{\nu_{\alpha}} (E_\nu)=\sum_{\alpha} \frac{N_{\nu_{\alpha}}}{4\pi \, d^ 2}\, 
\eta_{\nu_{\alpha}}^{\mathrm{SN}} (E_\nu),
\end{equation}
($\alpha = e, \mu, \tau$) where $N_{\nu_{\alpha}}$ is the number of (anti)neutrinos emitted from 
a supernova source at a typical distance (here we used $d = 8.5 \, \mathrm{kpc}$) and 
$\eta_{\nu_{\alpha}}^{\mathrm{SN}}$ denotes the energy distribution 
of the (anti)neutrino flavour $\alpha$ \cite{Giom-Vergados}. We assume that the emitted SN-neutrino energy spectra $\eta_{\nu_{\alpha}}^{\mathrm{SN}} (E_\nu)$ 
resemble Maxwell-Boltzmann distributions that depend on the temperature $T_{\nu_{\alpha}}$ of the (anti)neutrino flavour $\nu_\alpha$ ($\tilde{\nu}_\alpha$). By convoluting the integrated cross section $\sigma_{\lambda,\nu_\alpha}(E_\nu)$ with the neutrino 
distributions, the signal produced on a terrestrial detector may be simulated as
\begin{equation}
\sigma^{sign}_{\lambda,\nu_\alpha}(E_\nu)=\sigma_{\lambda,\nu_\alpha}(E_\nu)\,\eta_{\nu_\alpha}^{\mathrm{SN}}(E_\nu).
\label{eq.signal}
\end{equation}
\begin{table*}[ht]
\centering
\begin{tabular}{{c|ccccccccc}}
\hline \hline \\
$\nu_{\alpha}$  & $(A,Z)$ & $\langle\sigma_{\mathrm{tot}}\rangle$ & $\langle\sigma_{\mathrm{SM}}\rangle$ & $\langle\sigma_{\mathrm{NU}}\rangle$ & $\langle\sigma_{\mathrm{FP}}\rangle$ & $\langle\sigma_{\nu_{\alpha} \rightarrow \nu_{e}}\rangle$ & $\langle\sigma_{\nu_{\alpha} \rightarrow \nu_{\mu}}\rangle$ & $\langle\sigma_{\nu_{\alpha} \rightarrow \nu_{\tau}}\rangle$ &\\ 
\hline 
& $ ^{48}\mathrm{Ti}$ & $5.32$ & $5.15 $ & $1.20 \times 10^{-2}$ & $4.66$ & - & $6.07 \times 10^{-5}$ & $6.50 \times 10^{-1}$ &  \\[-1ex]
\raisebox{1.5ex}{ $\nu_{e}$ } 
& $ ^{27}\mathrm{Al}$ & $1.57$ & $1.50 $ & $3.83 \times 10^{-3}$ & $1.35$ & - & $1.95 \times 10^{-5}$ & $2.09 \times 10^{-1}$ & \\
\hline 
 & $ ^{48}\mathrm{Ti}$ & $19.6$ & $15.2 $ & $1.93 \times 10^{-2}$ & $14.2 $ & $1.80\times 10^{-4}$ & - & $5.36 $ &  \\[-1ex]
\raisebox{1.5ex}{ $\nu_{\mu}$ } 
& $ ^{27}\mathrm{Al}$ & $6.07$ & $ 4.61 $ & $6.42 \times 10^{-3}$ & $4.27 $ & $6.00 \times 10^{-5}$ & - & $1.78 $  & \\
 \hline \hline
\end{tabular}
\caption{ Flux averaged cross sections $\langle\sigma_{\lambda,\nu_{\alpha}}\rangle$ (in $10^{-40} \, cm^2$)
for various supernova neutrino spectra parametrized by Maxwell-Boltzmann distributions.}
\label{table2}
\end{table*}
%
\begin{figure}[H]
\begin{center}
\includegraphics[width=0.40\textwidth]{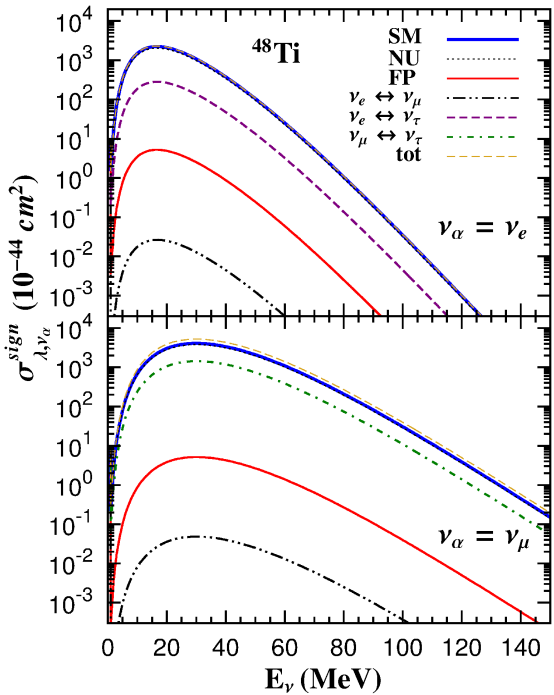}
\end{center}
\caption{ The convoluted cross sections, evaluated with Maxwell-Boltzmann distributions, that represent the 
expected signal to be recorded on $^{48}\mathrm{Ti}$ $\nu$-detector, $\sigma_{\lambda,\nu_\alpha}^{sign}(E_\nu)$. 
Due to the flavour dependence of the SN neutrino distribution, the energy-window of $\nu_{e}$ neutrinos 
signal is more narrow compared to those of $\nu_{\mu}$ and $\nu_{\tau}$ neutrinos.}\label{fig.3}
\end{figure}
%

The resulting signals, $\sigma^{sign}_{\lambda,\nu_\alpha}(E_\nu)$, obtained by inserting in Eq. 
(\ref{eq.signal}) the  cross sections $\sigma_{\lambda,\nu_{\alpha}}$, are plotted 
in Fig. \ref{fig.3}. Note that, in contrast to the original cross sections, now $\sigma^{sign}_{\nu_\alpha\to
\nu_\beta} \neq \sigma^{sign}_{\nu_\beta \rightarrow \nu_\alpha}$. 
Figure \ref{fig.3} shows that for incoming $\nu_\mu$ neutrinos the signal 
$\sigma^{sign}_{\lambda,\nu_\mu}$ presents an appreciably wider energy range compared to that of $\nu_e$ 
and that the maximum peak is shifted towards higher energies following the features of the distributions 
$\eta_{\nu_\alpha}^{\mathrm{SN}}(E_\nu)$.
The simulated cross sections of Fig. \ref{fig.3} reflect the characteristics of the incident neutrino spectrum
of a specific flavour $\alpha$ having its own position of the maximum peak and width of the distribution 
$\eta_{\nu_\alpha}^{\mathrm{SN}}$. We remind that, as usually, for incoming $\nu_e$ neutrinos, the distribution 
$(\eta_{\nu_e}^{\mathrm{SN}} + \eta_{\tilde{\nu}_e}^{\mathrm{SN}})/2$ is used. 

In SN neutrino simulations, another useful quantity is the flux averaged cross section 
\cite{Kosm-Oset} which in our notation is written as
\begin{equation}
\langle\sigma_{\lambda,\nu_\alpha}\rangle = \int\sigma_{\lambda,\nu_\alpha}(E_\nu)\,
\eta_{\nu_\alpha}^{\mathrm{SN}}(E_\nu) \, dE_\nu \, .
\end{equation}
The results for $\langle\sigma_{\lambda,\nu_\alpha}\rangle$, obtained by using our angle-integrated cross sections are listed in Table \ref{table2}. We note that our flux averaged cross sections
differ by about $30 \%$ from those of \cite{Barranco}.

\begin{figure}[H]
\begin{center}
\includegraphics[width=0.40\textwidth]{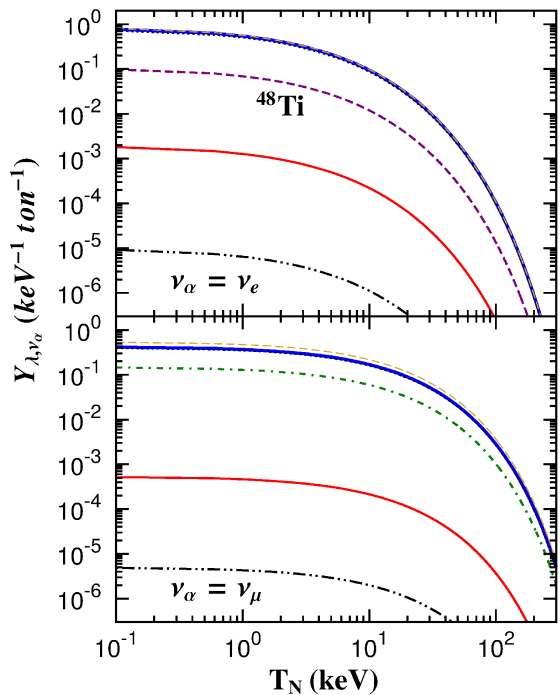}
\end{center}
\caption{ Differential event rate, $Y_{\lambda, \nu_{\alpha}}(T_N)$, as a function of the nuclear 
recoil energy, $T_N$, for $^{48} \mathrm{Ti}$ $\nu$-detector. The line labeling is same to that of 
Fig. \ref{fig.3}.} \label{fig.4}
\end{figure}

From experimental physics perspectives, it is also interesting to make predictions for the differential 
event rate of a $\nu$-detector \cite{Horowitz,Biassoni,vtsak-tsk-2}. The usual expression 
for computing the yield in events is based on the neutrino flux, $\Phi_{\nu_\alpha}$. 
To include the NSI of neutrinos with nuclei, the yield in events $Y_{\lambda,\nu_{\alpha}}(T_N)$, is \cite{Horowitz,Biassoni}
\begin{equation}
Y_{\lambda, \nu_\alpha}(T_N) = N_t \int \Phi_{\nu_\alpha} \, dE_{\nu} \int \frac{d\sigma_{\lambda,\nu_\alpha}}{d\cos\theta} \,   
\delta\left(T_N - \frac{q^2}{2 M}\right) \, d\cos \theta \, , \label{diff.rate}
\end{equation}
where $N_t$ is the total number of nuclei  in the detector material. Assuming a detector filled 
with one tone $^{48}$Ti, we evaluated differential event rates $Y_{\lambda,\nu_\alpha}(T_N)$ for several 
supernova scenarios. 
These results, are plotted in Fig. \ref{fig.4} where for each particular interaction, the corresponding 
neutrino flux has been considered. We see that, the respective results for the NU and FC processes, 
especially the case of $\nu_\mu \rightarrow \nu_e$ transition, present appreciably small contributions 
and that, the lower the energy recoil, the larger the potentially detected number of events. Hence, 
for the observation of non-standard $\nu$-nucleus events, detector medium with very low energy-recoil 
threshold is required. 

With the above results for $Y_{\lambda, \nu_{\alpha}}(T_N)$, one can obtain the total number of counts 
by integrating Eq. (\ref{diff.rate}) above the energy threshold, $T_N^{thres.}$, of the detector in 
question. For the $^{48}\mathrm{Ti}$ nucleus, assuming $T_N^{thres.} \approx 1\, \mathrm{keV}$, we find 
about $13.5$ events/ton for the SM process but only 
$10^{-3}$ events/ton for the flavour changing $\nu_\mu\leftrightarrow \nu_e$
reaction, i.e. about four orders of magnitude less events \cite{Pap-Kosm-NPA}. We also conclude that, 
for making accurate predictions of the total number of counts, the nuclear structure parameters play 
significant role. Thus, for the $\nu_\mu\rightarrow \nu_e$ transition we end up with about $29 \%$ 
less events, compared to those given by the approximation of Ref. \cite{Barranco}. On the other hand, 
adding up the total number of events for the three SM processes of the form, $\nu_\alpha \to \nu_\alpha$, 
we end up with only $2 \%$ less events than those provided from the formalism of Refs. \cite{Horowitz,Biassoni}. 

It is worth noting that, the choice of the target nucleus plays also a key role, since a light nuclear 
target may yield high energy recoil tails but less counts. On the contrary, a heavy nuclear 
target provides more counts and yields low-energy recoils making the detection more difficult. This leads 
to the conclusion that the best choice for a nuclear detector must consists of a combination of light 
and heavy nuclear isotopes \cite{Biassoni}. 
\subsection{New stringent limits  on $\epsilon_{\mu e}^{f V}$ from $\mu^-\rightarrow e^-$ conversion}
\label{section_NSI_limits}
In the last part of this analysis, we exploit our channel-by-channel cross sections calculations in order 
to provide new limits for the NSI parameters $\epsilon_{\mu e}^{f P}$, coming out of the present and future 
experimental constraints of cLFV $\mu^-\rightarrow e^-$ 
conversion as follows. The authors of Ref. \cite{Davidson} (assuming that cLFV arises from loop diagrams 
involving virtual W's) found that the couplings of charged leptons with quarks 
are given by $C \epsilon_{\alpha\beta}^{f P}$, where $C \approx 0.0027$. 
Consequently, for the $\nu_\mu\leftrightarrow\nu_e$ transition the NSI parameters are related with the experimental upper limits of $\mu^{-} \rightarrow e^{-}$ conversion as \cite{Davidson}
\begin{equation}
\epsilon_{\mu e}^{f P}= C^{-1} \sqrt{R_{\mu e}^{(A,Z)}}  \, .
\label{eps_mue}
\end{equation}
In our calculations, up to this point we used the value $\epsilon_{\mu e}^{f V} = 2.9 \times 10^{-4}$ 
resulting from the PSI upper limit, $R_{\mu e}^{Ti} < 6.1 \times 10^{-13}$ \cite{Wintz} (occasionally, 
this value is a more severe constraint compared to the value $\epsilon_{\mu e}^{f V} = 7.7 \times 10^{-4}$ 
used in \cite{Davidson} which came out of the upper limit $R_{\mu e}^{Ti} < 4.3 \times 10^{-12}$ \cite{Dohmen}).

Significantly lower upper limits on the NSI $\epsilon_{\mu e}^{f P}$ parameters of Eq. (\ref{R-lamb-alp}), 
are expected to be derived from the COMET, Mu2e, Mu2e at Project-X and PRIME/PRISM $\mu^-\to e^-$ conversion 
experiments. Then, one may compute new ratios $R_{\nu_\mu\leftrightarrow\nu_e}$ of the FC $\nu_e\leftrightarrow\nu_\mu$ 
reaction channel. The results for the NSI parameters $\epsilon_{\mu e}^{f V}$ and 
the respective ratios $R_{\nu_\mu \leftrightarrow \nu_e}$ are listed in Table \ref{table3}. 
\begin{table}[h]
\centering
\setlength{\tabcolsep}{0.5 em}
\begin{tabularx}{0.48\textwidth}{{c|cccc}}
\hline \hline
\\
 Parameter & COMET & Mu2e & Project-X  &  PRIME  \\
  \hline  
 $\epsilon_{\mu e}^{f V} \times 10^{-6} $ & $3.70$ & $2.87$ & $0.52$ &  $ 0.37$ \\
 $R_{\nu_{\mu} \leftrightarrow \nu_{e}} \times 10^{-10}$ & $21.2$ & $13.0$ & $0.42$ & $0.19$ \\
 \hline \hline
\end{tabularx}
\caption{Upper limits on the NSI parameters $\epsilon_{\mu e}^{f V}$ and the ratios 
$R_{\nu_\mu \leftrightarrow \nu_e}$ for the FC $\nu_\mu \leftrightarrow\nu_e$ reaction channel
resulting from the sensitivity of the $\mu^-\rightarrow e^-$ conversion experiments. }
\label{table3}
\end{table}

\begin{figure}[H]
\begin{center}
\includegraphics[width=0.40\textwidth]{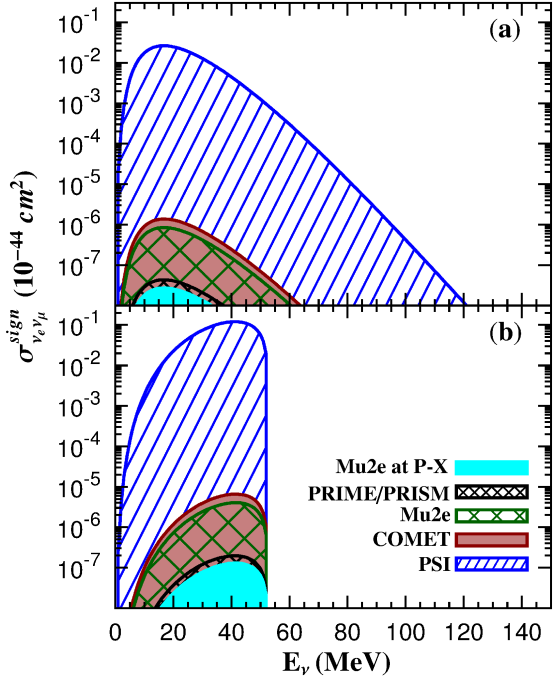}
\end{center}
\caption{Simulated $\nu$-signal, $\sigma^{sign}_{\nu_e \to\nu_\mu}$, of the FCNC process
$\nu_e + (A,Z)\to \nu_\mu + (A,Z)$ in $^{48}$Ti, for the PSI and PRIME/PRISM experiments 
and in $^{27}$Al, for the COMET, Mu2e and Mu2e at Project-X:
\textit{(a)} for supernova neutrinos and \textit{(b)} for pion-muon stopped neutrinos. 
The shaded area represents the excluded 
region of observation by the increased sensitivity of the designed experiments. 
For each plot the relevant NSI parameter $\epsilon_{\mu e}^{f P}$ of Table \ref{table3} has been employed.} 
\label{fig.5}
\end{figure}

Before closing we find interesting to plot the expected neutrino signals 
$\sigma^{sign}_{\nu_\mu\to\nu_e}(E_\nu)$ 
resulting by using the limits of Table \ref{table3} in two cases of $\nu$-spectra: 
(i) supernova neutrinos, and (ii) laboratory neutrinos
originating e.g from the BNB (Booster Neutrino Beamline) at Fermilab known as 
pion decay-at-rest (DAR) neutrinos \cite{pion-DAR-nu,Louis}.
In the first case the simulated cross sections are obtained by employing the Supernova $\nu$-spectra, $\eta_{\nu_{\alpha}}^{\mathrm{SN}}$, 
discussed before \cite{Horowitz,Biassoni} and the results are illustrated in Fig. \ref{fig.5}(a).
In the second case, the simulated cross sections are obtained by considering the laboratory 
neutrino distribution of the stopped pion-muon neutrinos produced 
according to the reactions $\pi^+ \to\mu^+ + \nu_\mu$, $\mu^+\to e^+ +\nu_e+\tilde{\nu}_\mu$
\cite{pion-DAR-nu,Louis}. 
In these experiments the emitted $\nu_{e}$ neutrino spectrum is described by the normalized 
distribution $\eta_{\nu_{\alpha}}^{lab.}$, $\alpha=e,\mu$ \cite{Kosm-Oset,vtsak-tsk-2}. The simulated laboratory neutrino signal $\sigma^{sign}_{\nu_{e} \rightarrow \nu_{\mu}}$ is shown 
in Fig. \ref{fig.5}(b). 

As can be seen, in both cases the exceedingly high sensitivity of the designed experiments reduces 
drastically (compare Figs. \ref{fig.3} and \ref{fig.5}) the area of observation of the $\nu$-signals $\sigma^{sign}_{\nu_e \to \nu_\mu}(E_\nu)$.
%

We should note that for models based on non-unitary lepton mixing matrix (including seesaw), constraints on $n_{\alpha \beta}$  (related to $\epsilon_{\alpha \beta}^{f P}$ within normalisation factors \cite{Forero}) may similarly come out. Obviously, for NSI considering both $d$ and $u$ quarks, $n_{\alpha \beta}$ enter the nuclear matrix elements of Eq. (\ref{GV}). 
\section{Conclusions}
In conclusion, we explored NC non-standard $\nu$-nucleus processes with realistic nuclear structure 
calculations. As a first step, we evaluated cross sections for the dominant coherent channel (incoming neutrino 
energies $0 \le E_\nu \le 150$ MeV, which include stopped pion-muon neutrinos, supernova neutrinos, etc). 
We have examined partial, integrated and total coherent cross sections and determined constraints for the ratios
$R_{\nu_\alpha \to \nu_\beta}$ of all relevant reaction channels with respect to the SM cross section. 
Furthermore, we provided results for the differential event rates and the total number of events 
assuming one ton of $^{48}$Ti as $\nu$-detector material. In view of operation of the muon-to-electron 
conversion experiments, searching for the exotic $\mu^-\rightarrow e^-$ conversion, we concentrated 
on the $^{48}$Ti nucleus previously used as stopping target by the PSI experiment and recently proposed 
to be used by the PRIME experiment at J-PARC. Similarly we have studied  the $^{27}$Al as $\nu$-detector, 
proposed to be used as muon stopping target in the sensitive Mu2e and COMET experiments. 

New stringent upper limits (up to even three orders of magnitude lower than those previously put) on 
the NSI (FC) parameters $\epsilon_{\mu e}^{f V}$ are extracted by using the experimental sensitivity 
of the $\mu^-\rightarrow e^-$ conversion experiments and our present results. By comparing our results with those of
other methods we concluded that the nuclear physics aspects (reflecting the reliability and accuracy 
of the cross sections), largely affect the coherent $gs \rightarrow gs$ transition rate, a result especially useful for supernova $\nu$-detection probes and low-energy laboratory neutrinos.   

Finally, we would like to remark that, $\mu^{-} \rightarrow e^{-}$ transition experiments at sensitivities
down to $10^{-16}-10^{-18}$ have excellent capabilities to search for evidence of new physics and to study its flavour structure. These well designed experiments at Fermilab and at J-PARC, could be the starting point 
of such a new effort, which would complement the neutrino programs. They have significant potential for constraining the NSI parameters and shed light on FCNC processes 
in the leptonic sector and specifically on the existence of the charged-lepton mixing.


\section{Acknowledgements} 
TSK wishes to thank Robert Bernstein and Graham Kribs for the financial support to attend the Project-X Physics Study 2012 and the warm hospitality he enjoyed at Fermilab.




\section{References}
\bibliographystyle{elsarticle-num}
\bibliography{<your-bib-database>}

\begin{thebibliography}{00}
\bibitem {Kuno} Y. Kuno, Y. Okada,  Rev. Mod. Phys. 73 (2001) 151.
\bibitem {Molzon} W.R. Molzon, Improved tests of muon and electron flavor symmetry 
         in muon processes, Springer Tracts Mod. Phys. 163 (2000) 105.
\bibitem {Bernstein-Cooper}  R.H. Bernstein, P.S. Cooper, Phys. Rept., in press, [hep-ex]/1307.5787.
\bibitem {COMET} Y.G. Cui et al. (COMET Collab.), KEK Report 2009-10; 
         COMET Collab., The COMET Proposal to JPARC, JPARC Proposal, 2007; 
         COMET Collab., A. Kurup, Nucl. Phys. Proc. Suppl. 218 (2011) 38. 
\bibitem {Mu2e-proposal} Proposal to Search for $\mu^- + N \rightarrow e^- + N$ with a Single 
         Event Sensitivity Below $10^{-16}$, (Mu2e Experiment) by the Mu2e Collab., Fermilab, 
         Oct. 10, 2008. 
\bibitem {Mu2e} R.J. Abrams et al. (Mu2e Collab. and Project), Mu2e Conceptual Design Report, 
         arXiv:1211.7019; Mu2e Collab., F. Cervelli, J. Phys. Conf. Ser. 335 (2011) 012073.

\bibitem {Bernstein} R. Bernstein, G. Kribs, \textit{Muon Physics at Project X}, summary talk Project X 
         Physics Study (PXPS12), Fermilab, June 14-28, 2012, Chicago USA.  
\bibitem {Kosm-talk} T.S. Kosmas, \textit{Nuclear Physics Aspects of the Exotic $\mu \rightarrow e$ Conversion 
         in Nuclei}, Invited talk in \cite{Bernstein}.  

\bibitem {Valle} J.W.F. Valle, Nucl. Phys. B (Proc. Suppl.) 229-232 (2012) 23.

\bibitem {Davidson} S. Davidson, C. Pena-Garay, N. Rius, A. Santamaria, JHEP 03 (2003) 011.
\bibitem {Barranco} J. Barranco, O.G. Miranda, T.I. Rashba, JHEP 0512 (2005) 021.
\bibitem {Amanik_2005} P.S. Amanik, G.M. Fuller, B. Grinstein, Astropart. Phys.  24 (2005) 160.
\bibitem {Tomas-Valle-10} A. Esteban-Pretel, R. Tomas, J.W.F. Valle, Phys. Rev. D 81 (2010) 063003.

\bibitem {Kosm_PhysRep} T.S. Kosmas, J.D. Vergados, Phys. Rep. 264 (1996) 25l.

\bibitem {Kos_Dep_Wal} F. Deppisch, T.S. Kosmas, J.W.F. Valle,  Nucl. Phys.  B 752 (2006) 80.

\bibitem {Wintz} P. Wintz, in Proceedings of the First International Symposium on Lepton and Baryon Number Violation (1998), edited by H.V. Klapdor-Kleingrothaus and I.V. Krivosheina (Institute of Physics Publishing, Bristol and Philadelphia), p.534. Unpublished.
\bibitem {Dohmen}  C. Dohmen et al. (SINDRUM-II Collab.), Phys. Lett. B 317 (1993) 631.

\bibitem {mu2e-px} K. Knoepfel et al., (2013), arXiv:1307.1168 [physics.ins-det].
\bibitem {PRIME} R.J. Barlow, Nucl. Phys. B Proc. Suppl. 218 (2011) 44.
\bibitem {Kuno-PRIME} Y. Kuno, Nucl. Phys. B Proc. Suppl. 225 (2012) 228.
\bibitem {MEG} J. Adam et al. (MEG Collab.), Phys. Rev. Lett. 110 (2013) 201801.
\bibitem {mu-eee} U. Bellgardt et al. (SINDRUM Collab.), Nucl. Phys. B 299 (1988) 1.

\bibitem {Schechter-Valle} J. Schechter, J.W.F. Valle, Phys. Rev. D 22 (1980) 2227; Phys. Rev. D 25 (1982) 774.
\bibitem {Forero} D.V. Forero, S. Morisi, M. Tortola, J.W.F. Valle, JHEP 1109 (2011) 142.
\bibitem {Dep-Valle} S.P Das, F.F. Deppisch, O. Kittel, J.W.F. Valle, Phys. Rev. D 86 (2012) 055006.
\bibitem {Amanik_2007} P.S. Amanik, G.M. Fuller, Phys. Rev. D 75 (2007) 083008.

\bibitem {Friedland-solar} A. Friedland, C. Lunardini, C. Pena-Garay, Phys. Lett. B 594 (2004) 347.
\bibitem {Friedland-atm2} A. Friedland, C. Lunardini, Phys. Rev. D 72 (2005) 053009.
\bibitem {Scholberg} K. Scholberg, Phys. Rev. D 73 (2006) 033005.

\bibitem {Barranco-Walle} J. Barranco, O.G. Miranda, C.A. Moura, J.W.F. Valle, Phys. Rev. D 73 (2006) 113001.
\bibitem {Mir-Tort-Walle} O.G. Miranda, M.A. Tortola, J.W.F. Valle, JHEP 0610 (2006) 008.
\bibitem {Bar-Mir-Rashba} J. Barranco, O.G. Miranda, T. I. Rashba, Phys. Rev. D 76 (2007) 073008. 

\bibitem {Kosm-A570} T.S. Kosmas, J.D. Vergados, O. Civitarese, A. Faessler, Nucl. Phys. A 570 (1994) 637.

\bibitem {Kos_Kov_Schm} T.S. Kosmas, S. Kovalenko, I. Schmidt,  Phys. Lett. B 511 (2001) 203; Phys. Lett. B 519 (2001) 78.

\bibitem {Kosm-Oset} T.S. Kosmas, E. Oset, Phys. Rev. C 53 (1996) 1409.


\bibitem {SK} Super-Kamiokande Collab., Y. Fukuda, et al., Phys. Rev. Lett. 86 (2001) 5651.
\bibitem {SNO} SNO Collab., Q.R. Ahmad, et al., Phys. Rev. Lett. 89 (2002) 011302.
\bibitem {KamLAND} KamLAND Collab., K. Eguchi, et al., Phys. Rev. Lett. 90 (2003) 021802.

\bibitem {Hirata-Bionta} K. Hirata, et al., Phys. Rev. Lett. 58 (1987) 1490;
         R.M. Bionta, et al., Phys. Rev. Lett. 58 (1987) 1494.
\bibitem {Keil} M.T. Keil, G.G. Raffelt and H.-T. Janka, Astrophys. J. 590 (2003) 971.

\bibitem {Haxton} W.C. Haxton, Phys. Rev. Lett. 60 (1988) 768.

\bibitem {pion-DAR-nu} A.A. Aguilar-Arevalo et al., Phys. Rev. D 79 (2009) 072002. 
\bibitem {Louis} W.C. Louis, Prog. Part. Nucl. Phys. D 63 (2009) 51.
\bibitem {Horowitz} C.J. Horowitz, K.J. Coakley, D.N. McKinsey, Phys. Rev. D 68 (2003) 023005. 
\bibitem {Biassoni} M. Biassoni, C. Martinez, Astropart. Phys. 36 (2012) 151.
\bibitem {Freedman} D.Z. Freedman, Phys. Rev. D 9 (1974) 1389 ; A. Drukier, L. Stodolsky, Phys. Rev. D 30 (1984) 2295.

\bibitem {Giom-Vergados} Y. Giomataris, J.D. Vergados, Phys. Lett. B 634 (2006) 23.
\bibitem {Monroe_Fischer} J. Monroe, P. Fisher, Phys. Rev. D 76 (2007) 033007; A.J. Anderson, J.M. Conrad, E. Figueroa-Feliciano, K. Scholberg, J. Spitz, Phys. Rev. D 84 (2011) 013008.

\bibitem {Don-Wal} T.W. Donnelly, J.D.Walecka, Nucl. Phys. A 274 (1976) 368.
\bibitem {vtsak-tsk-1} V. Tsakstara, T.S. Kosmas, Phys. Rev. C 83 (2011) 054612.
\bibitem {vtsak-tsk-2} V. Tsakstara and T.S. Kosmas, Phys. Rev. C 84 (2011) 064620.

\bibitem {deVries} H. De Vries, C.W. De Jager, C. De Vries, At. Data and Nucl. Data Tables 36 (1987) 495536.



\bibitem {Pap-Kosm-NPA} D.K. Papoulias, T.S. Kosmas, Nucl. Phys. A, to be submitted.
\end{thebibliography}

\end{document}